\documentclass[twocolumn]{aastex6}
\usepackage{multirow}

\shorttitle{The effects of individual metals}
\shortauthors{Beom et al.}

\begin{document}

\title{The effects of individual metal contents on isochrones 
\\for $C$, $N$, $O$, $N\MakeLowercase{a}$, $M\MakeLowercase{g}$, 
$A\MakeLowercase{l}$, $S\MakeLowercase{i}$, and $F\MakeLowercase{e}$}

\author{Minje Beom\altaffilmark{1}, Chongsam Na\altaffilmark{1}, 
Jason W. Ferguson\altaffilmark{2} and Y. -C. Kim\altaffilmark{1}}

\affil{\altaffilmark{1}Department of Astronomy, 
Yonsei University, Seoul 03722, Republic of Korea; yckim@yonsei.ac.kr\\
\altaffilmark{2}Department of Physics, Wichita State University, 
Wichita, KS 67260-0032, USA}

\begin{abstract}

The individual characteristics of C, N, O, Na, Mg, Al, Si, and Fe on isochrones 
have been investigated in this study. 
Stellar models have been constructed for various mixtures 
in which the content of each element is changed up to the extreme value 
reported in recent studies, and the changes in isochrone shape have been analyzed 
for the various mixtures. 
To express the abundance variation of different elements with a single parameter, 
we have focused on the relative changes in the total number of metal ions. 
A review of the shape changes revealed that Na, Mg, and Al 
work the same way in stellar models, 
as the well-known fact that C, N, and O 
have the same reactions in the stellar interior. 
In addition, it was found that in high-metallicity conditions 
the influence of Si and Fe on the red giant branch becomes smaller than 
that of Na, Mg, and Al closer to the tip.
Furthermore, the influence of Fe on the main sequence 
is larger than that of Na, Mg, Al, and even Si.

\end{abstract}


\section{Introduction}

In almost all Galactic globular clusters (GGC), 
the abundance variations of individual elements have been reported. 
Since the study of \citet{Sim68,Sim70},
the significance of the variation in C, N, and O abundances has been emphasized 
owing to their high abundances in chemical compositions 
\citep{Ren77,Dem80,Roo85}. 
Stellar models, in which chemical composition is adopted from observations, 
have been constructed for comparison studies (e.g., \citet{Van85}). 
Because ‘chemical anomalies’ of various other elements 
and anti-correlations of $CNONa$ and $MgAl$ in GGCs have been reported 
\citep{Osb71,Can98,Gra01,Coh05,Car09a,Car09b,Lee13}, 
it has become more important to consider observational results. 
For various other elements, 
the effects on stars of the abundance variations have been investigated 
(e.g., \citet{Dot07,Van12}). 
Furthermore, some studies suggest scenarios 
for explaining the general abundance pattern present in the stars of GGCs,
including chemical compositions with positive variations in C, N, Na, 
negative variations in O \citep{Sal06,Cas08,Pie09,Ven09}, and 
positive variations in Al and negative variations in Mg content \citep{Cas13}.

\begin{table*}[ht!]
\centering
\caption{Input Physics and Parameters}
\label{tbl1}
\begin{tabular}{cc}
  \hline
  \hline
  Input Physics & Source \\
  \hline
  \multirow{2}{*}{Solar mixture}	& \citet{GS98}\\
    & ($Z$/$X$ = 0.0230) \\
  Abundances and ratio of $\alpha$ elements & $[\alpha/Fe] = 0.3$ \citep{Van00} \\
  OPAL Rosseland mean opacities 	& \citet{OPALOPAC}\footnotemark[1] \\
  Low temperature opacities      & \citet{Fer05}\footnotemark[1] \\
  Equations of state				& OPAL EOS \citep{OPALEOS} \\
  Helium diffusion 				& \citet{Tho94} \\
  Mixing length parameter	    & ${\it l}/H_{p} = 1.859$ (solar calibrated value) \\
  \hline
\end{tabular}
\footnotetext[1]{The opacity tables for the mixtures have been generated for this work.}
\end{table*}

Spectroscopic data of some GCCs show substantially more variations 
in the contents of a few elements than those previously recognized. 
For example, abundance variations of Al above $1.0~dex$ in [Al/Fe] 
have been reported in the stars of NGC 2808 and NGC 6752. 
It has been generally considered that changes in the content of 
less abundant elements such as Na and Al cause 
only minor effects on stellar evolution (e.g. \citet{Van12}).
However, their abundance variations up to extreme values 
reported by recent observations
may have considerable impacts on stellar evolution.

We have chosen eight elements, including C, N, O, Na, Mg, Al, Si, and Fe, 
to investigate cases in which their extreme values are reached, 
based on the values reported in previous studies. 
For various contents of these individual metals, stellar models and isochrones 
are constructed to analyze their effects and physical changes.
In this study, we focus on the variation in total particle number of metal ions, 
and we present the changes and shifts of isochrones in a systematic way. 
The changes in the isochrone shape versus the relative number variation of metal ions 
are investigated and the results indicate that the behaviors of Na, Mg, and Al 
are similar in the stellar interior. 
Thus, these three elements can be considered as a group,
which is similar to the well-recognized group of C, N, and O. 
Moreover, the influence of Fe on isochrones differs 
from those of other metal elements. 

The details of the stellar model constructions are given in section 2. 
The method of presenting the isochrone characteristics for systematic comparison 
is discussed in section 3. 
Section 4 presents the results and analysis based on
the isochrone shape change in terms of the changes in the total numbers of metal ions. 
Further discussion and a summary are given in sections 5 and 6, respectively.

\section{Stellar Model Construction}

\subsection{Code and input physics}

Stellar evolutionary models were constructed 
by using a version of the Yale Stellar Rotational Evolution Code (YREC), 
which is a standard stellar evolutionary code utilized for $Y^{2}$ isochrones 
\citep{Yi01,Yi08,Kim02}. 
The models in this study are compatible with the $Y^{2}$ isochrone set, 
although two updates have been made in the computation. 
First, the reference solar mixture of \citet{GS98} was used 
rather than that of \citet{GN93} used for the $Y^{2}$ isochrone projects. 
Although recent studies on the solar mixture have been conducted 
\citep{Asp05,Asp09,Lod09,Caf11,Ste16}, 
studies of helioseismology and neutrinos prefer 
the higher metallicity of mixtures reported in older studies. 
Thus, the solar mixture of \citet{GS98} was chosen for this work.
As \citet{Van12} stated, the results in this type of differential study 
are not affected by the selection of the reference mixture.
Second, the OPAL equation of state (OPAL EOS) 
was updated to its later version \citep{OPALEOS}. 
The change associated with the two updates is less 
than the typical observational error \citep{Beom14}. 
Table \ref{tbl1} summarizes the input physics employed for the computations in this work. 

For clarity, the following two points are worth mentioning. 
First, the cosmic chemical enrichment, specifically $\Delta Y/ \Delta Z$, 
was considered in the computation of the reference isochrones. 
In the comparison isochrones, however, 
to focus on the effects of abundance changes of the individual elements, 
the He content was kept constant even though the total metal abundance $Z$ was changed. 
This treatment is discussed further in the following subsection. 
Second, the stellar models in this work were constructed assuming 
that the ratio between metal contents in a mixture does not change 
during the evolution of a star.
The surface chemical mixture, which is accessible through spectroscopic observations, 
is the integrated result of various physical mechanisms 
such as dredge-up, diffusion, and dynamics in the stellar atmosphere. 
However, our knowledge of these physical processes for each metal element 
is still rather limited. 
Because these physical processes have not been included, 
the ratio between metal contents in a mixture did not change 
during the evolution of the stellar models.

\subsection{Construction of stellar models and isochrones}

\begin{table}[t!]
\centering
\caption{The chemical abundance grids for the reference stellar models}
\label{tbl2}
\begin{tabular}{lccc}
  \hline
  \hline
  Globular Cluster & $Z$\footnotemark[1] & $Y$\footnotemark[2] & [Fe/H] \\
  \hline
  Metal poor	& 0.0002 & 0.2304 & -2.173\\
		& 0.0005 & 0.2310 & -1.775\\
  Intermediate	& 0.0010 & 0.2320 & -1.473\\
		& 0.0020 & 0.2340 & -1.170\\
		& 0.0040 & 0.2380 & -0.866\\
  Metal rich	& 0.0070 & 0.2440 & -0.618\\
  Extremely metal rich 	& 0.0120 & 0.2540 & -0.375 \\
  \hline
\end{tabular}
\footnotetext[1]{The mass fraction of the elements heavier than the helium}
\footnotetext[2]{The mass fraction of the helium}
\end{table}


\begin{table*}[ht!]
\centering
\caption{The observed values of $[m/Fe]$ for the metal elements}
\label{tbl3}
\begin{tabular}{ccccc}
  \hline
  \hline
  \multirow{2}{*}{Elements} & Number Fraction 
  & \multicolumn{2}{c}{$\Delta[m/Fe]$} & \multirow{2}{*}{References} \\
  & in the Reference Mixture & \multicolumn{2}{c}{in Observations} & \\
  \hline
  C  & 0.14824 & -0.5 & +0.5 & \citet{Car05}\\
  N  & 0.03724 & \multicolumn{2}{c}{+1.6} & \citet{Car05,Coh05}\\
  O  & 0.60388 & \multicolumn{2}{c}{-0.6} & \citet{Car09a}\footnotemark[1]\\
  Na & 0.00187 & -0.3 & +0.5 & \citet{Car09a}\footnotemark[1]\\
  Mg & 0.03396 & -0.8 & +0.2 & \citet{Lee13}\\
  Al & 0.00069 & \multicolumn{2}{c}{+1.5} & \citet{Car09b}\footnotemark[1]\\
  Si & 0.03243 & \multicolumn{2}{c}{+0.4} & \citet{Car09b}\footnotemark[1]\\
  Fe & 0.01416 & \multicolumn{2}{c}{+0.4} & \\
  \hline
\end{tabular}
\footnotetext[1]{\citet{Car09a,Car09b}, “Na-O anti-correlation and HB VII and VIII.”}
\end{table*}

The ranges of mass and metallicity for the stellar model construction 
were set for globular clusters in our Galaxy. 
The metallicity range of $[Fe/H] = -2.173 \sim -0.375$ ($Z = 0.0002 \sim 0.012$), 
which extends slightly beyond the usual range, was adopted. 
Including a cosmic primordial He abundance of 0.23 
and a $\Delta Y/\Delta Z$ value of two, 
the chemical compositions for the reference models are summarized in Table \ref{tbl2}. 
For isochrones of 9 Gyr and older with this metallicity, 
the mass range was chosen as $0.7\sim1.1~M_{\odot}$ with $0.05~M_{\odot}$ increments. 
Considering the globular clusters in our Galaxy, 
the mixture of \citet{GS98} with $\alpha$ elements enhanced by 0.3 dex 
was adopted as the reference mixture for this work. 
The ratios of $\alpha$ elements were taken from \citet{Van00}. 

\begin{table*}[ht!]
\centering
\caption{The list of the metal compositions for each cases}
\label{tbl4}
\begin{tabular}{lcccccccccc}
  \hline
  \hline
  \multirow{2}{*}{Case} & \multicolumn{2}{c}{Metallicity\footnotemark[1]} & \multicolumn{8}{c}{Abundance of Individual Elements ($\log$~$N_{m}$)\footnotemark[2]}  \\
    & $Z$\footnotemark[3] & [Fe/H] & C & N & O & Na & Mg & Al & Si & Fe \\
  \hline
  $Reference$ & 0.0010 & -1.473 & 7.166 & 6.566 & 7.776 & 5.266 & 6.526 & 4.836 & 6.506 & 6.146 \\ 
  $C-0.5$     & 0.0009 & -1.473 & 6.666 & $\cdots$ & $\cdots$ & $\cdots$ & $\cdots$ & $\cdots$ & $\cdots$ & $\cdots$ \\
  $C+0.5$     & 0.0012 & -1.473 & 7.666 & $\cdots$ & $\cdots$ & $\cdots$ & $\cdots$ & $\cdots$ & $\cdots$ & $\cdots$ \\
  $N+0.8$     & 0.0012 & -1.473 & $\cdots$ & 7.366 & $\cdots$ & $\cdots$ & $\cdots$ & $\cdots$ & $\cdots$ & $\cdots$ \\
  $N+1.6$     & 0.0022 & -1.472 & $\cdots$ & 8.166 & $\cdots$ & $\cdots$ & $\cdots$ & $\cdots$ & $\cdots$ & $\cdots$ \\
  $O-0.3$     & 0.0007 & -1.473 & $\cdots$ & $\cdots$ & 7.476 & $\cdots$ & $\cdots$ & $\cdots$ & $\cdots$ & $\cdots$ \\
  $O-0.6$     & 0.0006 & -1.473 & $\cdots$ & $\cdots$ & 7.176 & $\cdots$ & $\cdots$ & $\cdots$ & $\cdots$ & $\cdots$ \\
  $Na-0.3$    & 0.0010 & -1.473 & $\cdots$ & $\cdots$ & $\cdots$ & 4.966 & $\cdots$ & $\cdots$ & $\cdots$ & $\cdots$ \\
  $Na+0.5$    & 0.0010 & -1.473 & $\cdots$ & $\cdots$ & $\cdots$ & 5.766 & $\cdots$ & $\cdots$ & $\cdots$ & $\cdots$ \\
  $Mg-0.8$    & 0.0010 & -1.473 & $\cdots$ & $\cdots$ & $\cdots$ & $\cdots$ & 5.726 & $\cdots$ & $\cdots$ & $\cdots$ \\
  $Mg+0.2$    & 0.0010 & -1.473 & $\cdots$ & $\cdots$ & $\cdots$ & $\cdots$ & 6.726 & $\cdots$ & $\cdots$ & $\cdots$ \\
  $Al+0.8$    & 0.0010 & -1.473 & $\cdots$ & $\cdots$ & $\cdots$ & $\cdots$ & $\cdots$ & 5.636 & $\cdots$ & $\cdots$ \\
  $Al+1.5$    & 0.0010 & -1.473 & $\cdots$ & $\cdots$ & $\cdots$ & $\cdots$ & $\cdots$ & 6.336 & $\cdots$ & $\cdots$ \\
  $Si+0.4$    & 0.0011 & -1.473 & $\cdots$ & $\cdots$ & $\cdots$ & $\cdots$ & $\cdots$ & $\cdots$ & 6.906 & $\cdots$ \\
  $Fe+0.4$    & 0.0011 & -1.073 & $\cdots$ & $\cdots$ & $\cdots$ & $\cdots$ & $\cdots$ & $\cdots$ & $\cdots$ & 6.546 \\
  \hline
\end{tabular}
\footnotetext[1]{The metallicities for the cases of $Z_{ref.}=0.0010$ $([Fe/H]_{ref.}=-1.473)$.
                  The abundance variation is so slight 
                  that the metallicities seems to be unchanged for most of the cases.}
\footnotetext[2]{The abundances for the cases of $Z_{ref.}=0.0010$ on the scale $\log N_{He}=11.00$; 
                  $A_{m} = \log N_{m}/N_{He} +11.00$.}
\footnotetext[3]{The mass fraction for the heavier elements than He.}
\end{table*}

To focus on the effects of each element clearly, 
the abundance of each specific element was changed individually. 
We chose to fix all of the other elements except for H. 
For enhancement of a specific element, for example, 
the content of H is decreased by the increment of the target element. 
This causes the stellar model to have slightly higher $Z$ and $[Fe/H]$. 
However, any change shown in a model was confirmed beforehand 
to be attributed primarily to the abundance change of the target element 
rather than to the associated change in the $\Delta X/X$ value. 
Because the relative variation in $X$, i.e., $\Delta X/X$, is very small, 
the changes in the value of $[Fe/H]$ in this work are also very small, at less than 0.01. 
The net effect of the increased metal and the decreased H combined 
is an overall enhancement of opacity throughout the stellar interior, 
which causes the tracks to shift toward the lower temperature. 
Even though H is the most important opacity source in the stellar interior, 
the changes in the $\Delta X/X$ value are not large enough 
to compensate for the opacity change due to the increased metal content. 
The treatment of this work produces tracks comparable to those 
produced by \citet{Van12}, in which $[Fe/H]$ and $Y$ were fixed. 

The isochrones were generated by utilizing the method 
developed for the third version of $Y^{2}$ isochrone projects. 
Each track was divided into several parts 
based on the key equivalent evolutionary phases (EEP) 
in the evolution, the zero-age main sequence (ZAMS), 
the main-sequence turnoff point (MSTO), 
the base of the giant branch (the base of GB), 
the bumps on the GB, and the GB tip. 
In each part between the key EEPs, numerous secondary EEPs were assigned 
on the basis of the track length on the $\log$ $T_{eff}$ versus $\log$ $L$ diagram.
Isochrones were generated by interpolating tracks at the same secondary EEPs.

\subsection{Abundance variations}

The eight elements selected for this study, 
C, N, O, Na, Mg, Al, Si, and Fe, 
are either abundant metals in the stellar chemical makeup or 
elements having significant abundance variation 
according to recent observations of the stars in GGCs. 
Their abundance variations were taken from the extreme values, 
or the mean values of sub-population stars in the observed globular clusters. 
To compare the effects of Fe and Si directly, 
the abundance variation for Fe was chosen to be the same as that of Si. 
Table \ref{tbl3} shows the target elements, their variation, and the references. 
For N, O, and Al, of which the variations are quite large, 
the intermediate value cases were included. 
The various cases in this study are summarized in Table \ref{tbl4}. 
The first column in the table lists the cases, 
with the name consisting of the atomic symbol for the element changed 
and the amount of the change in $[m/Fe]$ with respect to the reference mixture. 
For example, $C-0.5$ indicates that the abundance of C is decreased by 0.5 dex in [C/Fe] 
with respect to the reference mixture. 
For example, the remaining columns of the table present the chemical composition 
set in the case of $Z_{ref.} = 0.001$, where the subscript $ref.$ refers to the reference model. 
For the second and the third columns, 
the metallicities of each case are expressed in $Z$ and $[Fe/H]$. 
As mentioned in section 2.2, the total metallicities are slightly modified as well. 
The last eight columns show the abundances by number for the eight target elements; 
these values were calculated on the scale of $\log N_{He}=11.00$.

\section{Analysis method: a systematic presentation of the changes in isochrone shape}

In previous studies, such as \citet{Van12} and \citet{Pie09}, 
the effects of abundance variation of individual elements 
have been discussed extensively. 
The abundance variations of an element 
mainly influence the opacity and energy generation, 
resulting in shape changes of evolutionary tracks and isochrones. 
To present these effects quantitatively, we examined the differences 
in the locations of MSTO and GB, and the length of sub-GB between isochrones. 

\begin{figure}[ht!]
\figurenum{1}
\includegraphics[width=\columnwidth]{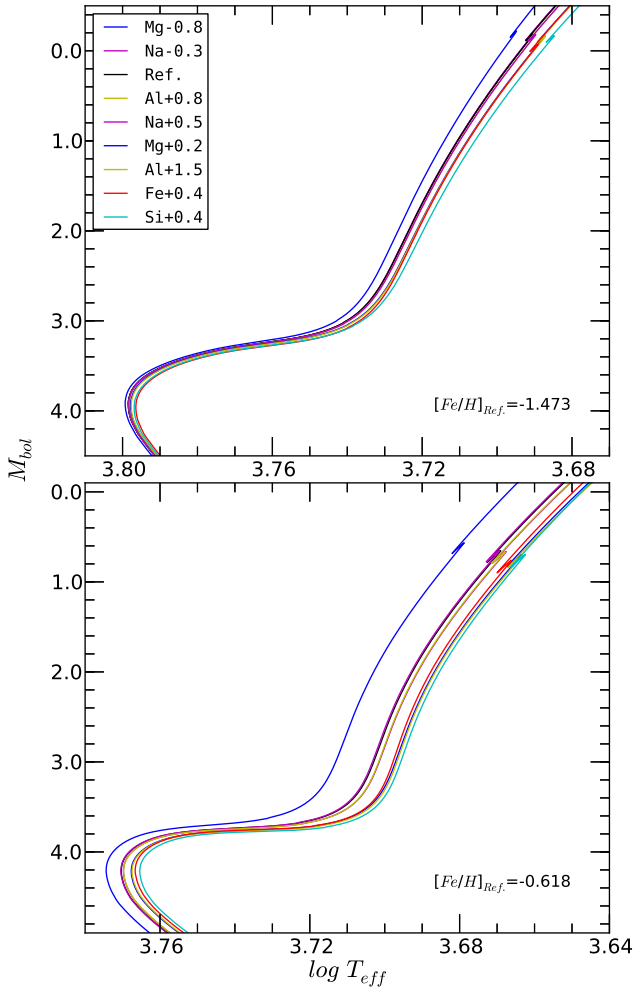}
\caption{12 Gyr isochrones for various $Na$, $Mg$, $Al$, $Si$, or $Fe$ abundances. 
 The colored lines present the 12 Gyr isochrones of the different mixtures 
 as indicated in the legend. The order of the mixture in the legend 
 was chosen according to the shift of the isochrones 
 with respect to the reference isochrone. 
 The spread of isochrones is significant only at the RGB 
 at low metallicity (upper panel), 
 whereas the spread is significant even at the MS and at the RGB 
 at high metallicity (bottom panel).}
\label{NaFe_iso}
\end{figure}

The net effect on isochrones is measured by considering an isochrone as a whole, 
from the main sequence (MS) to the red GB (RGB) tip. 
It is common knowledge that the influence of C, N, and O is mainly on the MS phase 
and that of Na, Mg, Al, Si, and Fe is mainly on the RGB. 
This simple description, however, is not sufficient. 
When the metallicity is sufficiently high in the MS stars 
to create effective temperatures of less than about 3.80 in 
$\log T_{eff}$, Na, Mg, Al, Si, or Fe 
can affect the opacity significantly in the outermost parts of these stars. 
As a result, a metal-rich star can be influenced by the effect of this high surface opacity 
during its entire lifetime. 
In Figure \ref{NaFe_iso}, this phenomenon is clearly indicated by 
a comparison of the upper and bottom panels, 
which show isochrones with metallicities of $[Fe/H] = -1.473$ and $-0.618$, respectively. 
The colored lines represent isochrones with various abundance variations. 
Contrary to that in the upper panel, 
the high-metallicity isochrones in the bottom panel show noticeable shifts of the MS 
that are comparable to those of the RGB. 
Moreover, such modifications on the MS phase are comparable to, or more than, 
those caused by C, N, or O. 
Therefore, the changes in the locations of both the MS and RGB 
were estimated for all elements considered. 
Because the effect of the abundance variation 
can differ depending on the locations along an RGB, 
the shifts in various locations were estimated from the RGB base to the tip.

In this study, the changes in the shapes of isochrones are presented in 
three parameters: the MS location, RGB location, and length of the sub-GB. 
In determining the difference in MS location between isochrones, 
the hottest points were selected as the MSTOs. 
The RGB locations were compared at four different levels set at points 
1.5, 3.0, 4.5, and 6.0 brighter in magnitude than the MSTO. 
The dimmest and the brightest levels among these locations 
represent the RGB base and tip, respectively. 
Except for the cases of C, N, and O variations, 
the location differences are considered only in terms of effective temperature 
rather than luminosity because the changes in the latter 
were small in the mass and metallicity ranges of this study. 
For the length of the sub-GB, the difference in effective temperature 
between the MSTO and the base of the RGB were estimated. 
The base of the RGB was set to the location 
1.5 mag brighter than that of the MSTO. 
It should be noted that the length of the sub-GB in this study 
is commonly used for age estimation of globular clusters 
\citep{Dem90,Sar90,Van13}.

\section{Results}
\begin{figure}[ht!]
\figurenum{2}
\includegraphics[width=\columnwidth]{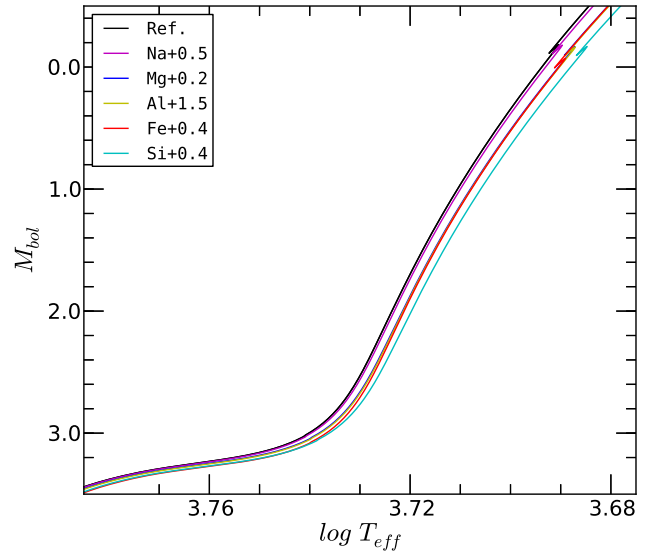}
\caption{From the sub-GB to RGB bump of the 12 Gyr isochrones for six cases are given, 
 namely, $reference$, $Na+0.5$, $Mg+0.2$, $Al+1.5$, $Si+0.4$, and $Fe+0.4$. 
 The order of the mixture in the legend was chosen 
 according to the shift of the isochrones 
 with respect to the reference isochrone. 
 The isochrones of $Mg+0.2$, $Al+1.5$, and $Fe+0.4$ show the almost the same shape 
 so that their lines nearly overlap.}
\label{Al_iso}
\end{figure}

In this section, we analyze the changes in isochrone shape 
for mixtures with abundance variations of individual elements. 
We focus on the parameter, namely, the relative changes in the total number of metal ions 
introduced by the variation of the individual elements. 
Because these changes are approximately the same for $Mg+0.2$, $Fe+0.4$, and $Al+1.5$, 
the isochrones of these three mixtures almost overlap in the Hertzsprung–Russell diagram (HRD). 
When the isochrone shape change is reviewed in terms of this parameter, 
it is found that Na, Mg, and Al have similar effects. 
It is widely accepted that C, N, and O can be considered as one group 
and that the total sum of their content can be utilized for describing
the effect on stellar structure and evolution 
(e.g. \citet{Sim70,Ren77,Dem80,Van85}). 
The newly identified group of Na, Mg, and Al in this paper 
can be considered similarly. 
In addition, Fe shows distinctive characteristics from other elements.
As that of Si, the influence of Fe is less than those of Na, Mg, and Al
near RGB tip, in metal rich stars.
In MS stars, however, the influence of Fe is much larger than those of Na, Mg, Al, and even Si.

\subsection{Influence of extreme variation in abundance}

It has been considered that the less abundant metals 
such as Al have limited influence on stellar evolution. 
However, huge increases or decreases reported by recent observations
may have considerable impacts.
For $Al+1.5$, which is the extreme case based on the values from observations, 
the isochrone shows noticeable changes with respect to the reference in Figure \ref{Al_iso}. 
The isochrone for $Al+1.5$ shows the same degree of change 
as those for $Mg+0.2$ and $Fe+0.4$. 
In fact, the three isochrones are difficult to distinguish in the figure. 
On the contrary, the isochrone for Na does not show significant change in the figure. 
Although Na has about twice the abundance of Al in the reference mixture, 
the variation of $Na+0.5$ is not as large as that of $Al+1.5$. 
The amount of Na increase that makes a noticeable change in isochrone shape
will be discussed in section 5.1.

For the three cases of $Al+1.5$, $Mg+0.2$, and $Fe+0.4$, 
whose isochrones overlap in Figure \ref{Al_iso}, 
the actual changes in the total number of metal ions are similar 
even though the abundance variations presented in dex 
differ significantly, namely +1.5, +0.2, +0.4. 
Thus, we focused on the relative changes in the total number of metal ions 
($\Delta N_{Z}/N_{Z,ref.}$) to describe the abundance variation 
in the study of the effect of individual elements on isochrones.

The unit of dex ($[m/Fe]$) is generally used to express abundance variation. 
However, when comparing cases involving different elements, 
values expressed in $[m/Fe]$ can be misleading 
because it is a value scaled for the solar mixture. 
The mixture with enhanced C by 0.5, for example, 
shows more variation in the number of metal ions than that shown by 
the mixture with enhanced N by 0.8 
because C is more abundant than N in the solar mixture. 
For this reason, the abundances of abundant elements 
such as O, Mg, and Si, have mainly been considered 
in the chemical composition of stars \citep{Roo85,Van12}. 
Another issue is that the value is log scaled. 
For example, a comparison of mixtures with N enhanced by 1.6 and that by 0.8 in dex 
revealed a twofold difference in number 
although the actual change in the number of N particles differed about fourfold. 
Thus, to express the abundance variation using a single parameter, 
the relative changes in the total number of metal ions is examined in this study.

\begin{table}[t!]
\centering
\caption{The relative variation in total number of metal ions according to each case}
\label{tbl5}
\begin{tabular}{lr}
  \hline
  \hline
  \multirow{2}{*}{Mixture} & $\Delta N_{Z}/N_{Z,ref.}$\footnotemark[1] \\ 
    & \multicolumn{1}{c}{(\%)} \\ 
  \hline
  $Reference$ &   0.00   \\ 
  $C-0.5$     & -10.14   \\
  $C+0.5$     & +32.05   \\
  $N+0.8$     & +19.77   \\
  $N+1.6$     & +144.50  \\
  $O-0.3$     & -30.12   \\
  $O-0.6$     & -45.22   \\
  $Na-0.3$    & -0.09    \\
  $Na+0.5$    & +0.40    \\
  $Mg-0.8$    & -2.86    \\
  $Mg+0.2$    & +1.99    \\
  $Al+0.8$    & +0.37    \\
  $Al+1.5$    & +2.12    \\
  $Si+0.4$    & +4.90    \\
  $Fe+0.4$    & +2.14    \\
  \hline
\end{tabular}
\footnotetext[1]{The relative variation in total number of metal ions, expressed in percentage. Furthemore this parameter is independent of the total metallicity.}
\end{table}

\subsection{Abundance variation and net effect}

\begin{figure*}
\centering
\figurenum{3}
\includegraphics[width=120mm]{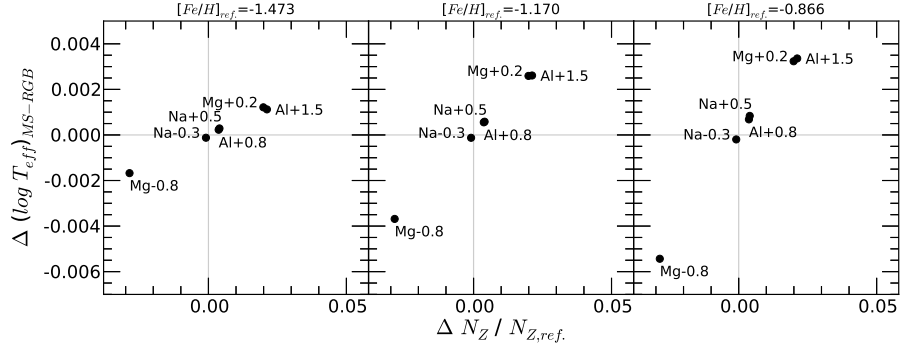}
\caption{For isochrones of various $Na$, $Mg$, or $Al$ contents,
 the number variation of metal elements versus the degree of net effect
 represented by the changes in the length of the sub-GB,
 at three different metallicity conditions is given.
 The x axis is the number variation of metal ions,
 whereas the y axis is the length of the sub-GB.
 The seemingly linear relation may indicate that $Na$, $Mg$, and $Al$
 have similar behavior in the stellar structure.}
\label{len_SGB_NaMgAl}
\end{figure*}

\begin{figure*}
\centering
\figurenum{4}
\includegraphics[width=120mm]{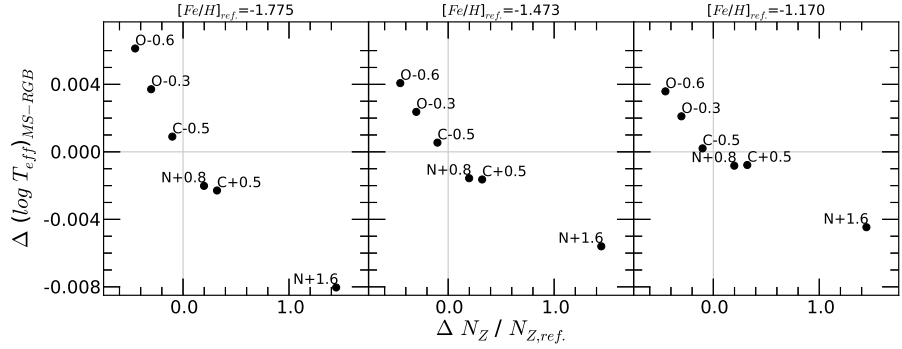}
\caption{Same as Figure \ref{len_SGB_NaMgAl}
except for isochrones of various $C$, $N$, and $O$ contents.
 Linearity can be found in each panel,
 which means $C$, $N$, and $O$ have similar behavior in stellar models.}
\label{len_SGB_CNO}
\end{figure*}

\begin{figure*}
\centering
\figurenum{5}
\includegraphics[width=120mm]{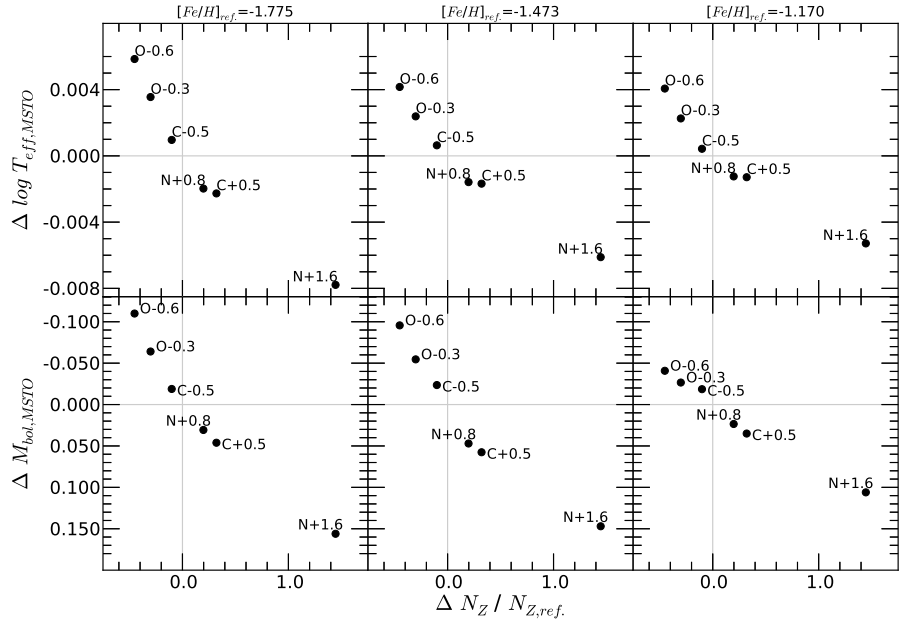}
\caption{For isochrones of various $C$, $N$, or $O$ contents,
 the number variation of metal elements versus the degree of net effect
 represented by the shifts of the MSTO is shown
 for three different metallicity conditions.
 The x axis is the number variation of metal ions,
 and the y axis is the shift of the MSTO on the HRD.
 A linear relation can be seen in all panels, as is shown in Figure \ref{len_SGB_CNO}.}
\label{l_MSTO_CNO}
\end{figure*}

\begin{figure}
\figurenum{6}
\includegraphics[width=\columnwidth]{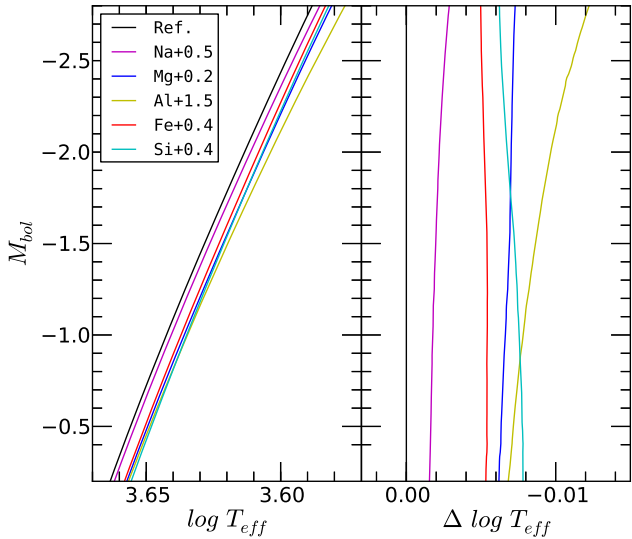}
\caption{Upper RGB of isochrones for various $Na$, $Mg$, $Al$, $Si$, and $Fe$ mixtures. 
 The colored lines represent the 12 Gyr isochrones of the different mixtures 
 as indicated in the legend. 
 The order of the mixture in the legend was chosen 
 according to the shift of the isochrones 
 with respect to the reference isochrone. 
 Upward along the RGB, the isochrones of $Si$ and $Fe$ 
 approach the reference isochrones, 
 which means that the influence of $Si$ and $Fe$ 
 decreases more than that of the other elements as the effective temperature decreases.}
\label{SiFe1}
\end{figure}

\begin{figure*}
 \centering 
 \figurenum{7}
 \includegraphics[width=155mm]{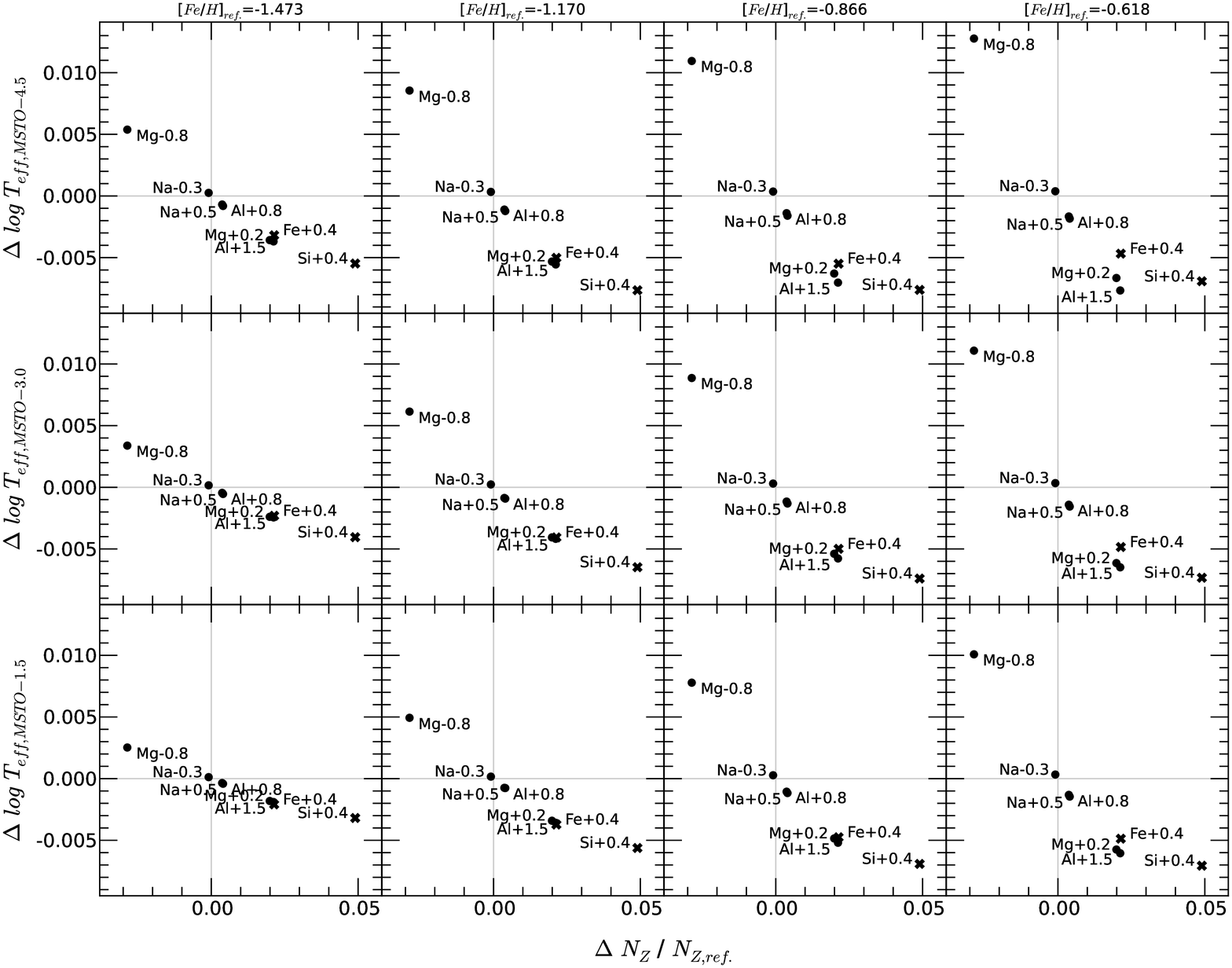}
 \caption{
 For isochrones of various $Na$, $Mg$, and $Al$ mixtures, 
 the number variation of metal elements versus the degree of net effect 
 represented by the shifts of the RGB is given for three different metallicity conditions. 
 The x axis is the number variation of metal ions, 
 and the y axis presents the shifts of the RGB measured at the various locations on the HRD. 
 The influence of $Si$ and $Fe$ is reduced more 
 than that of other elements as the metallicity increases 
 and the upper location of the RGB is approached.}
\label{l_RGB_NaFe}
\end{figure*}

To express the abundance variation of various elements 
in the same parameter domain, 
the chosen parameter is the relative change in particle number 
with respect to the reference mixture. 
Because its value is relative to the reference mixture, 
this parameter is independent of the total metallicity. 
For example, when the parameter is +0.3, 
the total number of metal ions in the comparison model 
is enhanced by $+30~\%$ over that in the reference model. 
The parameter is presented as 
\begin{eqnarray}
 \Delta N_{Z}/N_{Z,ref.} \nonumber &=& \frac{N_{Z,mod.}-N_{Z,ref.}}{N_{Z,ref.}} \nonumber \\
 \label{eq:def}
\end{eqnarray}
where $N_{Z}$ is the abundance in number for all metal elements, 
and the subscripts $ref.$ and $mod.$  indicate the stellar models 
before and after the modification, respectively. 
The parameter can be calculated in the following equation 
with variation in dex and the number fraction in the reference mixture:
\begin{eqnarray}
 \Delta N_{Z}/N_{Z,ref.} \nonumber &=& f_{m,ref.} (10^{\Delta[m/Fe]} - 1) \nonumber \\
 \label{eq:cal}
\end{eqnarray}
where $m$ is a target element having a change in abundance, 
and $f_{m,ref.}$ and $\Delta[m/Fe]$ are the number fraction in the reference mixture 
and the variation in $dex$, respectively; 
the other variables are the same as those in Equation~(\ref{eq:def}). 
The values of the parameter for the cases in this work are summarized in Table \ref{tbl5}. 
When the abundance change is expressed with the parameter, 
the degree of the shape change of the isochrones 
appears to be linearly related to the abundance variation.

\subsection{Grouping of elements and Their abundance sum}

The impact of abundance variation in Na, Mg, and Al contents 
is shown in Figure \ref{len_SGB_NaMgAl}. 
The figure presents the difference in sub-GB length for three different metallicities 
in terms of the abundance variation, 
which was introduced in the previous section.
In the figure, the points appear to form a straight line. 
This indicates that the changes in isochrone shape are linearly correlated with 
the relative changes in the number of metal ions 
regardless of the choice of element among Na, Mg, or Al. 
That is, when the number variations are the same, 
the isochrones for the three elements have similar shapes.
The same is true for the shifts of MS and RGB, 
which is discussed further in the following subsection.

This correlation can be described as behavior 
that the three elements affect similarly in stellar interior models.
The abundance change for these elements affects 
the interior models mainly through the opacity. 
As reported by \citet{Van12}, 
the abundance change of these elements increases the opacity 
in the similar temperature ranges of the stellar interior. 
These elements have similar atomic structures; 
therefore, their ionization potential is similar, 
and they donate a similar number of electrons in a certain physical condition. 
Therefore, the opacity is similarly changed 
when the abundance changes of these elements are the same. 
The effect on the energy generation, however, is very limited. 
Unlike C, N, and O, the impact of the elements on the energy generation is through 
only changes of the physical condition associated with the structural changes.

It is widely accepted that C, N, and O have similar roles 
and influence in a star and that the degree of the effects 
depends on $[C+N+O/Fe]$. 
Figure \ref{len_SGB_CNO} is the same as Figure \ref{len_SGB_NaMgAl} except for the elements. 
As expected, a linear relation can be detected in each panel 
even though the linearity appears to be weaker in Figure 4. 
This may have occurred because these elements 
are related to nucleosynthesis as well as opacity, 
whereas the group of Na, Mg, and Al is directly related only to opacity.
Figure \ref{l_MSTO_CNO} plots the shifts of MSTO in both effective temperature (upper panels) 
and brightness (lower panels). 
In all panels, a linear relation can be found. 
This again means that the degree of the net effect is the same 
when the number of the target element is changed by the same amount 
regardless of which element is chosen among C, N, and O.

In summary, it appears that the Na, Mg, and Al 
can be considered as one group. 
Similar to $[C+N+O/Fe]$, $[Na+Mg+Al/Fe]$ determines 
the degree of the effect on stellar models. 
Therefore, no significant change can be expected 
in stars in which the sum of Na, Mg and Al contents is not significantly changed. 
For example, in stars of the AGB phase the $Mg-Al$ reaction 
reduces the Mg and increases the Al contents. 
Thus, next-generation stars polluted by the winds from AGB stars 
show $Mg-Al$ anti-correlation. 
Because the sum of Mg and Al contents is not significantly changed, 
however, the effect associated with the abundance variation 
in these contents may be very limited.

\subsection{Distinct characteristics of Si and Fe \\from other elements}

In this section, we focus on the characteristics of Fe and Si 
that distinguish them from the other elements considered in this work. 
It is considered that Si and Fe generally behave in a manner 
similar to other metals, especially Mg, in stellar models (e.g., \citet{Van12}). 
In particular, Si has been considered to be similar to Mg 
owing to their roles in stellar models and their high abundance. 
However, \citet{Dot07} reported that 
the influence of Si on opacity becomes smaller than that of Mg at low temperatures. 
In the present study, the effect of Fe, as well as Si, was shown 
to be lower than that of other metals at low temperatures. 
In addition, the influence of Fe is much larger than those of Na, Mg, Al, and even Si.

In high-metallicity conditions, the influence of Fe as well as Si 
decrease near the RGB tip where effective temperature is lower. 
Figure \ref{SiFe1} shows the RGB of the 12 Gyr isochrones with $[Fe/H] = −0.866$. 
In the right panel, unlike others, the isochrones of $Si+0.4$ and $Fe+0.4$ 
are inclined toward the reference isochrone as they approach the RGB tip. 
The same is shown in Figure \ref{l_RGB_NaFe}, which also shows 
the location differences of the RGBs with respect to the abundance variation. 
The four columns of the panels represent the four different values of metallicity, 
and the three rows represent cases with different locations of RGB shift measurement. 
In the left panels of the figure, the points of Si and Fe are located near the line 
formed by the points of Na, Mg, and Al. 
In the upper and right panels, however, 
which represent the shift of RGB measured at the brighter points along RGB 
in addition to higher metallicity, the two points 
appear not to be associated with the linear relation formed by the other points. 
For the abundance variations of $Si+0.4$ and $Fe+0.4$, 
the shift of RGB was expected to be increased with the points located farther below. 
The fact that this did not occur can be interpreted 
as the lesser influence of Si and Fe in stars with low effective temperature, 
i.e., for higher metallicity and positions closer to the RGB tip. 
Furthermore, this effect is stronger for Si than Fe.
The influence of each element on opacity differs
in various temperature and density according to its atomic structure. 
Thus, it is expected that the influence of Na, Mg, and Al 
might differ from each other only in cases with higher metallicity. 
However, the metallicity is beyond the usual metallicity range 
of globular clusters in our Galaxy (above about $[Fe/H] = -0.5$). 
In comparison, the reduced influence of Si and Fe can be seen in the RGB stars 
of a metal-rich cluster above $[Fe/H] = -1.0$.

\begin{figure*}
 \centering 
 \figurenum{8}
 \includegraphics[width=120mm]{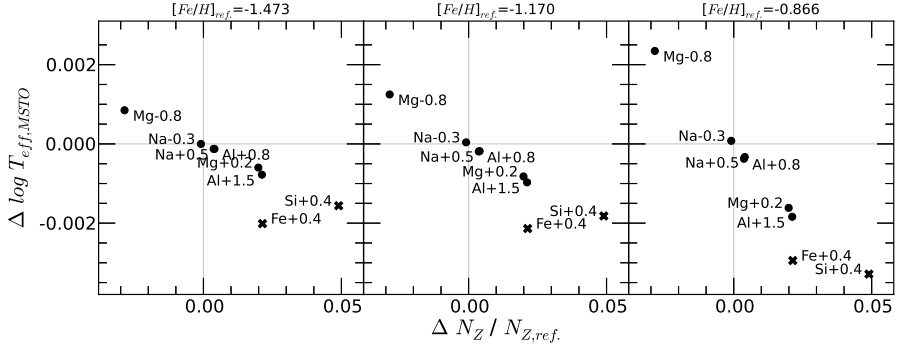}
 \caption{Same as the upper panels of Figure \ref{l_MSTO_CNO},
 except for isochrones of various $Na$, $Mg$, $Al$, $Si$ and $Fe$ cases. 
 The influence of $Fe$ near the MSTO is noticeably higher 
 than that of the other elements. 
 However, $Si$ appears to follow the trend of $Na$, $Mg$, and $Al$.}
\label{l_MSTO_NaFe}
\end{figure*}

\begin{figure*}[ht!]
 \centering
 \figurenum{9}
 \includegraphics[width=120mm]{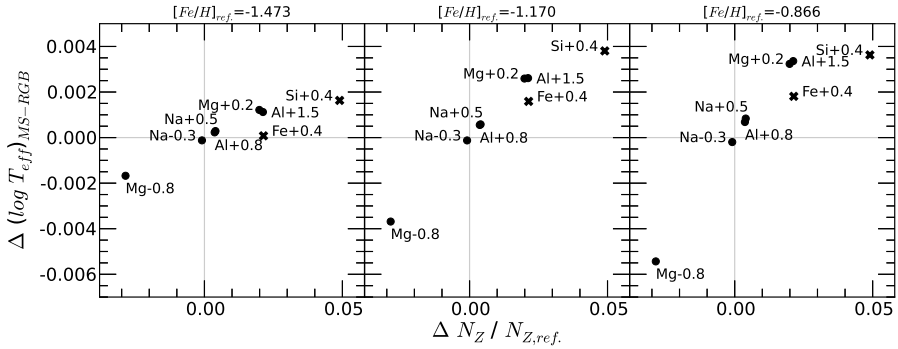}
 \caption{Same as Figure \ref{len_SGB_NaMgAl}, but including points of $Si+0.4$ and $Fe+0.4$.
 It should be noted that $Si+0.4$ and $Fe+0.4$ show less change
 than $Mg+0.2$ and $Al+1.5$ in the length of the sub-GB.}
\label{len_SGB_NaFe}
\end{figure*}

The influence of Fe near the MSTO is greater than 
that of Na, Mg, Al, and even Si. 
Figure \ref{l_MSTO_NaFe} shows the shift of the MSTO as a function of the abundance variation. 
The point of $Fe+0.4$ is located far below the line formed by other points. 
This indicates that more changes of the MSTO location than 
that of the other elements is expacted when the degree of the abundance variation is the same. 
Although the abundance variation of $Fe+0.4$ is similar to that of $Mg+0.2$ and $Al+1.5$, 
the MSTO of the isochrone shift of $Fe+0.4$ is twice that of the others. 
On the contrary, the point of $Si+0.4$ appears to be on the line 
formed by the points of the cases for Na, Mg, and Al. 
In the inner part of a stellar model, the opacity peak 
associated with Fe is at a different temperature range from that of the other elements; 
the opacity peaks of Na, Mg, Al, and Si occur almost in the same region \citep{Van12}.

These characteristics are shown in Figure \ref{len_SGB_NaFe}, 
which is the same as Figure \ref{len_SGB_NaMgAl} except for 
the addition of Fe and Si cases. 
In all panels of the figure, the points of $Fe+04$ and $Si+0.4$ 
do not appear to be associated with the linear relation of Na, Mg, and Al. 
The length of the sub-GB for $Fe+04$ and $Si+0.4$ is modified less 
because of the lower influence of Fe and Si on the RGB. 
Additionally, $Fe+04$ shows a significant shift of the MSTO, 
which is in the same direction as the RGB; thus, 
the change in the length of the sub-GB is less than that in other cases.

\section{Discussion}
The first subsection discusses 
abundance variations of less abundant metals.
The previous results show that their extreme variations 
can cause considerable change in isochrone shape.
The lower limits of abundance variations are discussed
based on the abundance variations and associated shape changes in the isochrone.
Furthermore, the results of the section 4 indicate that 
the two parameters, specifically $[C+N+O/Fe]$ and $[Na+Mg+Al/Fe]$
are key parameters in stars whose mixtures have $CNONa$ and $MgAl$ anti-correlations.
Thus, the last subsection discusses 
the implication of the new group of Na, Mg, and Al
in evolutionary point of view.

\subsection{Lower limits of abundance variation}

To determine the limiting value in the abundance variation 
that causes noticeable changes in the isochrones, 
the length of the sub-GB was investigated. 
It was assumed that the typical error of the age estimation 
utilizing the length of the sub-GB for a globular cluster is about $0.5~Gyr$ 
\citep{Van13}. 
For C and N, the enhancement of about $0.5~dex$ in $[C/Fe]$ or about $1.0~dex$ in $[N/Fe]$ 
cause a noticeable increase in the age estimation. 
Because of their low abundances in the reference mixture, however, 
any amount of depletion in the abundance of these elements 
cannot produce isochrones that resemble those that are $0.5~Gyr$ younger. 
Conversely, both enhancement and depletion are attainable for O.

For Na and Al, this factor depends on the metallicity. 
Metallicity lower than $[Fe/H] = -1.0$, $[Na/Fe] = 1.3~dex$ and above 
produce an isochrone appearing $0.5~Gyr$ younger, 
which corresponds with the enhancements of $\Delta[Na/Fe] = +1.0$ 
with respect to the reference mixture. 
For Al, the value is $[Al/Fe] = 1.2~dex$ and above, 
which is the same as $\Delta[Al/Fe] = +1.5$ with respect to reference mixture. 
For a metallicity higher than $[Fe/H] = -1.0$, 
the value is about $[Na/Fe] = 1.1~dex$ and above, 
which corresponds to $\Delta[Na/Fe] = +0.8$. 
And the value is $[Al/Fe] = 0.9~dex$ and above,
which is $\Delta[Al/Fe] = +1.2$. 
Because of the low abundance an isochrone resembling $0.5~Gyr$ older 
cannot be generated by any amount of depletion in Na and Al contents. 
For Mg, however, both enhancement and depletion may be considered.

As discussed in section 4.3, for elements in a group, 
the variation in the total sum rather than in individual variations is an important parameter. 
For $\Delta[C+N+O/Fe]$, an abundance variation of about $\Delta[C+N+O/Fe]^{+0.15}_{-0.20}$ 
and greater can produce recognizable changes in age estimation. 
Depending on metallicity, the variations beyond $\Delta[Na+Mg+Al/Fe]^{+0.20}_{-0.35}$ 
in the lower metallicity of $[Fe/H] = -1.0$ 
and $\Delta[Na+Mg+Al/Fe]^{+0.11}_{-0.14}$  
in the higher metallicity of $[Fe/H] = -1.0$ are considerable.

\subsection{Implication of the general abundance pattern}

The fact that Na, Mg, and Al can be considered as a group, 
such as the case of $CNO$, has an important implication 
on the studies of the GGCs that show $CNONa$ and $MgAl$ anti-correlations. 
According to the anti-correlations, 
the contents of C and N of second-generation stars is enhanced,
whereas that of O is depleted. 
Similarly, the abundance variation of Na and Al is opposite that of Mg. 
When the changes in the content of an individual element in a group cancel out, 
the resulting effect may be less than expected otherwise. 
For example, some second-generation stars have a mixture 
enhanced by $0.8~dex$ of Na and $1.0~dex$ of Al and depleted by 0.3 Mg \citep{Car09b}. 
The total abundance change is -0.0008 in $\Delta N_{Z}/N_{Z,ref}$ and -0.010 
in terms of $\Delta[Na+Mg+Al/Fe]$.
Therefore, no significant change is expected in the shape of its isochrone. 
This expectation, similar to that reported by \citet{Cas13}, 
can serve as an example of the case mentioned in the last part of section 4.3. 
For the case of \citet{Sal06}, 
second-generation stars have a mixture of 0.6 dex increase of C, 
$1.8~dex$ increase of N, and $0.8~dex$ decrease of O. 
The actual changes in total are 2.24 in $\Delta N_{Z}/N_{Z,ref}$ 
and 0.585 in $\Delta[C+N+O/Fe]$. 
Because of the huge increase in N, the total variation is very large 
even after some cancellation. 
Therefore, the isochrones for these second-generation stars 
differ significantly from those for first-generation stars.

For an isochrone shape on the theoretical HRD,
O is commonly taken as the representative among $CNO$ 
because it is the most abundant element. 
Similarly, Mg can represent Na, Mg, and Al. 
Thus, instead of all six elements of C, N, O, Na, Mg, and Al, 
only two elements of O and Mg need to be considered.
For the first example, the isochrone for $\Delta[Mg/Fe] = -0.010$ 
is the same as that with variations in the content of all three elements; 
the case of $\Delta[O/Fe] = 0.674$ is the same as that in the second example.

\section{Summary}

The abundance variations of individual elements 
have been reported in stars of GGCs; 
their effects on stellar evolution and isochrones 
have been investigated for extreme cases. 
In this study, stellar models and isochrones were constructed 
with mass and metallicity ranges for GGCs of 
$[Fe/H] = -2.173 \sim -0.375$ ($Z = 0.0002 \sim 0.012$) and $0.7 \sim 1.1 M_{\odot}$. 

In the analysis of its effect of an individual element, 
the changes in the total number of metal ions 
introduced by the abundance variation is shown to be a key parameter. 
The changes of the isochrone shape is expressed 
in terms of the change in the total number of metal ions. 
Analysis of the isochrones for various elements revealed that 
a few elements have similar roles in stellar models. 
In particular, Na, Mg, and Al can be considered as one group 
as in the case of C, N, and O. 
Therefore, the abundance variations can be considered 
as variations of $[Na+Mg+Al/Fe]$ and $[C+N+O/Fe]$.

The influence of Si and Fe on the MS and RGB 
differed from those of Na, Al, and Mg. 
Specifically, the influence of Si and Fe on the RGB became smaller 
than that of Na, Mg, and Al in a star with low effective temperature. 
Thus, the influence of Si and Fe decreases 
as the metallicity increases and in points ascending along the RGB. 
Additionally, the influence of Fe on the MS is larger 
than that of Na, Mg, Al, and even Si.

To produce noticeable change in isochrones, 
the variation in the content of elements should be larger 
than certain values. 
Depending on the metallicity, these values are about $[Na/Fe] = 1.1~dex$ or more, 
$[Al/Fe] = 0.9~dex$ or more with the enhancements of about $0.5~dex$ in $[C/Fe]$, 
and about $1.0~dex$ in $[N/Fe]$. 
This can also be presented as $\Delta[C+N+O/Fe]^{+0.15}_{-0.20}$ 
and $\Delta[Na+Mg+Al/Fe]^{+0.11}_{-0.14}$.

According to the general abundance pattern of second-generation stars, 
the abundances of the relatively less abundant elements 
such as C, N, Na, and Al were highly enhanced 
while those of the abundant elements such as O and Mg were depleted 
(e.g., \citet{Car09a,Car09b}). 
However, because Na, Mg, and Al can be considered 
as a single group similar to the case of C, N, and O, 
the abundance sum of $NaMgAl$ can also be utilized in the studies 
examining the effect caused by their abundance variation in stellar models.

\acknowledgments
This research is supported by Basic Science Research Program 
through the National Research Foundation of Korea (NRF) 
funded by the Ministry of Education, 
Science and Technology (NRF-2011-0025296). 
We thank Y.~-W. Lee for the initial suggestion 
and helpful discussions during this research.\\

\clearpage

\end{document}